\documentclass[amssymb,prb,twocolumn,showpacs]{revtex4}
\usepackage{epsfig}
\usepackage{dcolumn}
\usepackage{amsmath}
\begin{document}
%\documentclass[12 pt,a4paper]{article} %selecciona el tipo de documento
%\usepackage[english]{babel} %selecciona el idioma
%\frenchspacing %trata los espacios despues de los puntos igual que los otros
%\usepackage{epsfig}
%\usepackage{amsmath}
%\usepackage[a4paper,dvips]{geometry}
%\geometry{textwidth=16 cm, textheight=22 cm}
%\begin{document}

%\title[Author guidelines for IOP Publishing journals in  \LaTeXe]
\title{Radiation-induced resistance oscillations in a 2D hole gas: a demonstration of a universal effect.}
%\author{J. I\~narrea and Gloria Platero}

\author{Jes\'us I\~narrea$^{1,2}$ and Gloria Platero$^2$}

\address{$^1$Escuela Polit\'ecnica
Superior,Universidad Carlos III,Leganes,Madrid,28911 ,Spain\\
$^2$Unidad Asociada al Instituto de Ciencia de Materiales, CSIC,
Cantoblanco,Madrid,28049,Spain.}

%\affiliation{Escuela Polit\'ecnica
%Superior, Universidad Carlos III, Leganes, Madrid, 28911, Spain}

%\date{\today}
%%%%%%%%%%%%%%%%%%%%%%%%%%%%%%%%%%%%%%%%%%%%%%%%%%%%%%%%%%%%%%%%%%%%%%%%%%%%%%
%\section{Abstract}
\begin{abstract}
We report on a theoretical insight about the microwave-induced
resistance oscillations and zero resistance states when dealing with p-type semiconductors and holes
instead of electrons.
We consider a high-mobility two-dimensional hole gas hosted in a pure Ge/SiGe quantum
well. Similarly to electrons we obtain radiation-induce resistance oscillations
and zero resistance states. We
analytically deduce a universal expression for the irradiated magnetoresistance,
explaining the origin of the minima positions and their $1/4$ cycle phase shift.
The outcome is that these phenomena  are universal and only depend on
 radiation and cyclotron frequencies.
We also study the possibility of having simultaneously two different
carriers driven by radiation: light and heavy holes. As a result
the calculated magnetoresistance reveals an interference profile due
to the different effective masses of the two types of carriers.

% In the same
%scenario we obtain two different resonance peaks at low enough temperature.
%We study the dependence on microwave power and temperature   obtaining
%a similar behaviour as with electrons.

\end{abstract}
%%%%%%%%%%%%%%%%%%%%%%%%%%%%%%%%%%%%%%%%%%%%%%%%%%%%%%%%%%%%%%%%%%%%%%%%%%%%%%
\maketitle
\section{ Introduction}
High-mobility two-dimensional electron systems (2DES)  are fantastic platforms for
studying transport  and coupling with
different  potentials,  static or time-dependent (radiation) in nano-systems.
In the last decade two of the most striking effects   involving radiation-matter coupling in
2DES were discovered:
the microwave-induced resistance oscillations (MIRO) and zero resistance states (ZRS)\cite{mani1,zudov1}.
They are indeed remarkable effects that surprised condensed matter community when
they were discovered.   Mainly because
they involve simultaneously radiation-matter interaction\cite{ina1} and
transport excited by radiation in a nanoscopic system.  On the other hand their discovery was considered also
very important, specially in the case
of zero resistance states, because they were obtained without
quantization in the Hall resistance. The interest in both effects is
 focussed not only on the basic
explanation of a physical effect but also on its potential
applications.
They are obtained
when 2DES, in high
mobility samples at low temperature ($\sim 1K$), are subjected to a perpendicular magnetic field ($B$) and
radiation (microwave (MW) band) simultaneously. In these experiments, for an increasing
radiation power ($P$), one first obtains longitudinal
magnetoresistance ($R_{xx}$) oscillations
  which turn
into zero resistance states (ZRS) at high enough $P$.

Many experiments have been carried out\cite{mani2,
mani3,willett,mani4,smet,yuan,mani5,wiedmann1,wiedmann2,kons1,kons2,vk,mani6,mani7} and theoretical explanations \cite{ina2,girvin,dietel,lei,rivera,shi,chepe0}
have been given to try to explain their physical origin,  but  to date,  it  still remains controversial. On the other hand, although these effects
have been thoroughly studied, they have been mainly based  on GaAs/AlGaAs quantum wells and hardly ever other materials
have been used. Yet, we can cite
interesting experimental results about radiation-induced magnetotransport oscillations on
a different non-degenerate 2D system such as electrons on liquid helium surface\cite{kons1,kons2}; they may share
a similar physical origin as MIRO. In this way we wonder if MIRO and ZRS are universal effects and if, as a result,
they can be observed in different platforms. For instance, different materials and
carriers such as holes working with valence bands in p-type materials\cite{zudovholes}. In contrast
to 2DES, a two-dimensional hole gas (2DHG) presents more non-linearities and a more
interesting dynamics when it comes to coupling with radiation.

In this article, we
demonstrate that these effects are universal phenomena and that
they can be obtained as well in a 2DHG.
%For this purpose we have improved  the radiation-driven electron orbits model
%developed by the authors
Based on previous results\cite{ina2,ina3,kerner,park}  we obtain a universal expression for irradiated $R_{xx}$.
According to it,  MIRO only  depend on
radiation and cyclotron frequencies and not on the type of semiconductor material.
We are able to explain the experimentally obtained MIRO extrema and node positions; the $1/4$ cycle phase shift
of MIRO minima.  We have applied the
results to the case of holes obtaining MIRO and ZRS in a high-mobility 2DHG hosted in a pure Ge/SiGe
quantum well.  According to this
theory, when a Hall bar is illuminated, the
orbit centers of the Landau states perform a classical trajectory consisting in a harmonic
motion along the direction of the current. Thus, the  2D carriers move in phase and
harmonically  at the radiation frequency altering dramatically the scattering
conditions and giving rise eventually to MIRO and, at higher $P$, ZRS.

Working with the valence band gives us a new scenario as  is the possibility of having
two different carriers sustaining the current and  being coupled simultaneously with radiation.
We expect that this situation will deeply change the MIRO profile.
They would be light and heavy holes being driven by MW and  giving rise to a clear interference regime being evidenced in the calculated $R_{xx}$.
In the same way, the interplay of lower temperatures ($T$) and higher $P$ can reveal two different resonance
peaks at different $B$ in $R_{xx}$, each one corresponding to heavy and light holes.
Finally, we have studied the hole-based MIRO dependence on $T$ and $P$ obtaining similar results as
with electrons.
For instance, the calculated dependence on $P$ follows  a sublinear power law
that has been already obtained in  previous experiments \cite{mani6} with electrons and theoretically
confirmed\cite{ina4,ina5}. As expected the corresponding exponent of the power law is around $0.5$.

\begin{figure}
\centering \epsfxsize=3.2in \epsfysize=5.9in
\epsffile{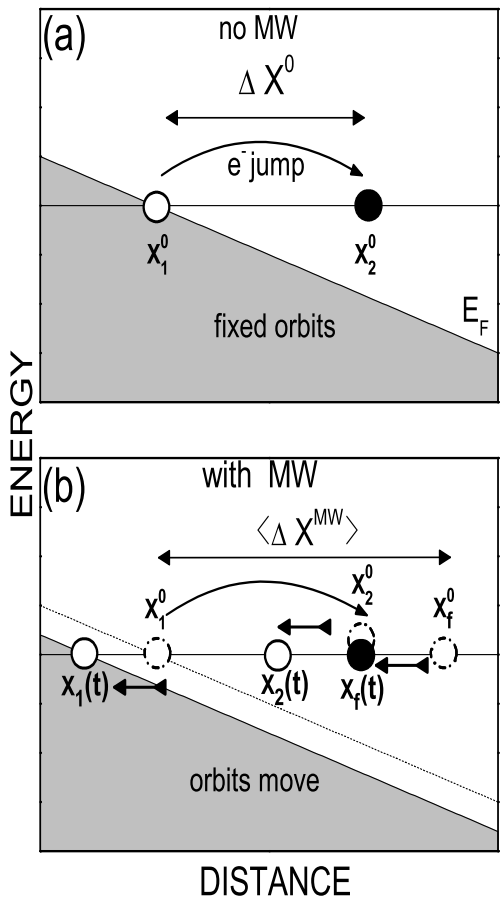}
\caption{Schematic diagrams for magnetotransport without and
with MW. In (a) no MW field is present and due to impurity scattering (elastic),
carriers jump between fixed-position orbits or Landau states. The positions
are fixed by the center of the respective orbits,  $X_{1}^{0}$  and $X_{2}^{0}$. The
average advanced distance is given by the difference of
the orbits center positions,  $\Delta X^{0}=X_{2}^{0}-X_{1}^{0}$.
In (b) orbits move backwards during the scattering jump and
on average carriers advance further than in the no-MW case.
This corresponds to a MIRO peak (see text below).
Now the average advanced distance is given by $\langle\Delta X^{MW}\rangle$.
}
\end{figure}

\section{ Theoretical Model}
The {\it radiation-driven electron orbits model}, was developed  to explain
the magnetoresistance response to radiation of a 2DEG at low $B$\cite{ina2,ina3,kerner,park}. We first obtain
an exact expression of the electronic wave vector for a 2DES in a
perpendicular $B$ and radiation.
 Thus, the total hamiltonian $H$ can be written as:
\begin{eqnarray}
H&=&\frac{P_{x}^{2}}{2m^{*}}+\frac{1}{2}m^{*}w_{c}^{2}(x-X)^{2}-eE_{dc}X +\nonumber \\
 & &+\frac{1}{2}m^{*}\frac{E_{dc}^{2}}{B^{2}}\nonumber-eE_{0}\cos wt (x-X) -\nonumber \\
 & &-eE_{0}\cos wt X \nonumber\\
 &=&H_{1}-eE_{0}\cos wt X
\end{eqnarray}
 $X$ is the center of the orbit for the electron spiral motion and
dependent on $B$ and $E_{dc}$ which is the DC
electric field in the $x$ direction.
%\begin{equation}
%X=\frac{\hbar k_{y}}{eB}- \frac{eE_{dc}}{m^{*}w_{c}^{2}}
%\end{equation}
$E_{0}$ is the intensity for the MW field and $H_{1}$ is the hamiltonian
corresponding to a forced harmonic oscillator whose orbit is
centered at $X$. $H_{1}$ can be solved exactly \cite{kerner,park}
making possible
 an exact solution for the total
wave function of $H$\cite{ina2,ina3,kerner,park,ina4}:
\begin{eqnarray}
\Psi_{N}(x,t)\propto\phi_{n}(x-X-x_{cl}(t),t)
\end{eqnarray}
where $\phi_{n}$ is the solution for the
Schr\"{o}dinger equation of the unforced quantum harmonic
oscillator. $x_{cl}(t)$ is the classical solution of a forced  and damped harmonic
oscillator:
\begin{equation}
x_{cl}(t)=\frac{e E_{o}}{m^{*}\sqrt{(w_{c}^{2}-w^{2})^{2}+\gamma^{4}}}\cos wt=A\cos wt
\end{equation}
%\begin{eqnarray}
%&&\Psi(x,t)=\phi_{n}(x-X-x_{cl}(t),t)\nonumber  \\
%&&\times  exp \left[i\frac{m^{*}}{\hbar}\frac{dx_{cl}(t)}{dt}[x-x_{cl}(t)]+
%\frac{i}{\hbar}\int_{0}^{t} {\it L} dt'\right]\nonumber  \\
%&&\times\sum_{m=-\infty}^{\infty} J_{m}\left[\frac{eE_{0}}{\hbar}
%X\left(\frac{1}{w}+\frac{w}{\sqrt{(w_{c}^{2}-w^{2})^{2}+\gamma^{4}}}\right)\right]
%e^{imwt}
%\end{eqnarray}
where $\gamma$ is a phenomenologically-introduced damping factor
for the  interaction of carriers with acoustic phonons.
$w$ is the  MW angular frequency where
$w=2\pi f$, $f$ being the frequency.
Since this model can be applied equally either to electrons or
holes, we will refer to them as {\it carriers} for the rest of the paper.

Then, the obtained wave
function is the same as the standard harmonic oscillator where the
center is displaced by $x_{cl}(t)$.
Thus, the carriers  orbit centers are not
fixed, but they oscillate harmonically at the MW frequency.
 This $radiation-driven$ behavior will dramatically affect  the
charged impurity scattering and eventually the conductivity. Next,
 we introduce the scattering suffered by the carriers due to
charged impurities.
 If the scattering is weak, we
can apply time dependent first order perturbation theory. First, we
calculate the impurity scattering rate, $W_{I}$,
between two $oscillating$ Landau states, the initial $\Psi_{n}(t)$ and the
final state $\Psi_{m}(t)$ \cite{ina2,ina3,ridley}.
%$W_{I}=1/\tau$
%being $\tau$ the scattering time.
%\begin{equation}
%W_{n,m}=\lim_{\alpha\rightarrow 0} \frac{d}{d t} \left|
%\frac{1}{i \hbar} \int_{-\infty}^{t^{'}}<\Psi_{m}(x,t) |V_{s}|\Psi_{n}(x,t)>e^{\alpha t}d t\right|^{2}
%\end{equation}
%where $V_{s}$ is the scattering potential for charged impurities\cite{4567ando}.
%\begin{equation}
%V_{s}= \sum_{q}\frac{e^{2}}{2 S \epsilon (q+q_{0})} \cdot e^{i
%\overrightarrow{q}\cdot\overrightarrow{r}}
%\end{equation}
%$S$ being the surface of the sample, $\epsilon$ the GaAs dielectric
%constant, and $q_{0}$ is the Thomas-Fermi screening
%constant\cite{ando}.

Secondly, and in order to calculate the
drift velocity, we find the average effective distance advanced by the carrier
in every scattering jump\cite{ina2,ina3,ridley}, $\Delta X^{MW}$.
Without radiation, one carrier in an initial Landau state $\Psi_{1}$
in an orbit center position $X_{1}^{0}$, undergoes a scattering process
and jumps to a final Landau state $\Psi_{2}$ with an orbit center position  $X_{2}^{0}$.
On average the carrier orbits center moves in the x direction a distance
given by the difference of the two orbits center positions, $\Delta X^{0}=X_{2}^{0}-X_{1}^{0}$  (see Fig. 1(a)).
With radiation, the carriers orbit center position  changes in the  $x$ direction harmonically with
time and is given according to our model by $X^{MW}(t)=X^{0}+A \cos (wt-\theta)$. $\theta$
is a general phase being calculated applying the initial conditions, i.e., for $t=0$,
 $X^{MW}(0)=X^{0}$, then $\theta= \pi/2 \Rightarrow$ $X^{MW}(t)=X^{0}+A \sin wt$.  In other words,
due to the MW field all the carrier orbit centers oscillate in phase back and forth
in the x direction with $A \sin wt$. On the other hand, after the MW is on, in a
given time  the carrier will undergo a scattering event. If this happens
when the orbits, driven by MW, move backwards we have the situation depicted in
Fig. 1(b). This corresponds to an increase in the average distance
advanced by the carriers giving rise to a MIRO peak.
If the orbits move forwards, we will have the opposite situation, a drop in the
average advanced distance,
producing a MIRO valley.

If at the moment of scattering the carrier is in the {\it oscillating} Landau state $\Psi_{1}(t)$ at the position $X_{1}(t)=X_{1}^{0}+A \sin wt$,
after a time $\tau$, what we call {\it flight time}, it will reach a {\it final oscillating}   Landau state $\Psi_{f}(t+\tau)$ located in
the position given by $X_{f}(t+\tau)=X_{f}^{0}+A \sin w(t+ \tau)$. In general this position
is not longer occupied by $\Psi_{2}$ since its position, $X_{2}(t)$, has been shifted a certain $\tau$-dependent distance and it is now
being taken by $\Psi_{f}$.
Then, the
advanced distance  under MW is:
\begin{eqnarray}
  \Delta X^{MW}&=&X_{f}(t+\tau)-X_{1}(t)\nonumber\\
  &=&X_{f}^{0}+A \sin w(t+\tau)-X_{1}^{0}-A \sin wt\nonumber\\
\end{eqnarray}
The  {\it flight time} $\tau$, is strictly the time it
takes the carrier to scatter from one orbit to another. This
time is part of the scattering time, $\tau_{s}$,  that is normally defined as
 the average time between scattering events. Therefore
 the scattering time would be made up of the time flight plus
 the time the carrier lies in the new orbit till the next
 scattering event takes place.

Now considering a stationary regime, i.e., averaging in time, we
obtain:
\begin{equation}
 \langle \Delta X^{MW}\rangle =X_{f}^{0}-X_{1}^{0}
\end{equation}
%From this expression we can conclude that   in the stationary regime the average
%advanced distance is calculated in terms of the mid-positions
%of the involved Landau states or orbits.
We can express $ \langle \Delta X^{MW}\rangle $ with respect to the
average advanced distance in the dark $\Delta X^{0}$:
\begin{equation}
 \langle \Delta X^{MW}\rangle =\Delta X^{0}+(X_{f}^{0}-X_{2}^{0})
\end{equation}
where the {\it distance shift}, $(X_{f}^{0}-X_{2}^{0})$, is going to be
the term responsible of the MW driven $R_{xx}$ oscillations or MIRO.
Therefore, according to that expression if $X_{f}^{0}>X_{2}^{0}$ we will have
a larger advanced distance in the $x$ direction producing a peak
in the conductivity and in turn in $R_{xx}$. In the opposite situation
if $X_{f}^{0}<X_{2}^{0}$ we would obtain a valley in $R_{xx}$ with respect to
the dark scenario.
When $X_{f}^{0}=X_{2}^{0}$ we would obtain a node, where $R_{xx}$ with radiation
is equal to the dark $R_{xx}$,  and  finally the most striking situation happens when
$X_{f}^{0}<X_{1}^{0}$ where ZRS occur.

To calculate the  {\it distance shift} we have to take into account, as
we said above, that the position
occupied by the orbit $\Psi_{2}$ at a given time $t$ will be taken by
the orbit  $\Psi_{f}$ after a time $t+\tau$: $X_{2}(t)= X_{f}(t+\tau)$.
On the other hand, mid-positions of the orbits are obtained when
%If we apply this condition to the mid-position of both orbits, i.e.,
when $wt=2\pi n$, $n$ being a positive integer, and substituting
this condition in the last equation: $X_{2}^{0}= X_{f}^{0}+A \sin w\tau$
%$\Rightarrow$
%\begin{equation}
%  X_{2}^{0}+A \sin wt=X_{f}^{0}+A \sin w(t+\tau)
%\end{equation}
%If we solve this equation for the mid-positions  positive number,
 We finally obtain an expression for the average  {\it distance shift}:
\begin{equation}
(X_{f}^{0}-X_{2}^{0})=-A \sin w\tau
\end{equation}

Substituting in Eq. 6 we can write for $ \langle \Delta X^{MW}\rangle $:
\begin{eqnarray}
 \langle \Delta X^{MW}\rangle &=&\Delta X^{0}-A \sin w\tau
% &=&\Delta X^{0}+A \cos (w\tau-\pi/2)
\end{eqnarray}
Thus in principle,  the two key variables to observe MIRO and ZRS are   $w$ and
 $\tau$. For instance, if for a fixed $w$, $\tau$ is very small then
$\sin w\tau\rightarrow 0$, then, there will be no effect on $R_{xx}$ and we would
obtain no MIRO. On the other hand, if $\tau$ is much larger than $T$, ($\tau >> T=\frac{2\pi}{w}$),
$T$ being the period of MW, then the average value of $\sin w\tau$ would be zero and no MIRO
will be observed either.
 Therefore we can conclude that in order to observe MIRO,
$\tau$ has to be of the order of the period of MW, otherwise the effect would vanish.

The longitudinal conductivity
$\sigma_{xx}$ is given by\cite{ridley}
\begin{equation}
\sigma_{xx}\propto \int dE \frac{\langle\Delta X^{MW}\rangle}{\tau_{s}}
\end{equation}
being $E$
the energy.
To obtain $R_{xx}$ we use
the relation
$R_{xx}=\frac{\sigma_{xx}}{\sigma_{xx}^{2}+\sigma_{xy}^{2}}
\simeq\frac{\sigma_{xx}}{\sigma_{xy}^{2}}$, where
$\sigma_{xy}\simeq\frac{p_{i} e}{B}$, $p_{i}$ being the holes density, and $\sigma_{xx}\ll\sigma_{xy}$.
Thus,
\begin{equation}
R_{xx}\propto - A  \sin w\tau
% \frac{eE_{o}}{m^{*}\sqrt{(w_{c}^{2}-w^{2})^{2}+\gamma^{4}}}\sin w\tau
\end{equation}
From this expression the MIRO minima positions fulfill the condition:
\begin{equation}
w\tau = \frac{\pi}{2}+2\pi n \Rightarrow w = \frac{2 \pi}{\tau}\left(\frac{1}{4}+ n \right)
\end{equation}
$n$ being a positive integer.
And for the MIRO maxima:
\begin{equation}
w\tau = \frac{3 \pi}{2}+2\pi n \Rightarrow w = \frac{2 \pi}{\tau}\left(\frac{3}{4}+ n \right)
\end{equation}

We can obtain also expressions for the MIRO nodes, or the points
where the radiation curve crosses the dark curve. If we take as a reference any MIRO peak,
the right node fulfills, $w = \frac{2 \pi}{\tau}( n-1/2)$. And the left one, $w = \frac{2 \pi}{\tau} n$.
To have an evaluation and a deeper physical meaning  of the {\it flight time} $\tau$ we can first
use a quantum mechanical approach using the time-energy uncertainty relation\cite{cohen}: $\Delta t\cdot \Delta E \simeq h$.
In our case, $\tau$  is the time it takes the
carrier to evolve from
the initial Landau state $\Psi_{n}$ to the final one  $\Psi_{m}$, so the uncertainty of time
is represented by $\tau$: $\Delta t =\tau$. During this time (flight between orbits), the state function of the
carrier can be assumed as a linear superposition of the two Landau states involved in the process,
and with respective energies: $E_{n}$   and $E_{m}$. If we measured the energy we would obtain
either $E_{n}$   or $E_{m}$, then the uncertainty of the energy is: $\Delta E = |E_{n}-E_{m}|= \hbar w_{c}$.
For the last expression we have assumed that $m=n+1$. This is a reasonable assumption
because if the carrier jumps from the Landau state with index $n$, the closest in energy  Landau state,
with index $n+1$, is the most likely state for the carrier to end up after the scattering event.
Therefore the uncertainty relation reads:
\begin{equation}
 \Delta t\cdot \Delta E = \tau \cdot \hbar w_{c} \simeq h\Rightarrow \tau \simeq\frac{2\pi}{w_{c}}=T_{c}
\end{equation}
where $T_{c}$ is the cyclotron period: the carrier {\it flight time} $\tau$, turns out
to be approximately equal to the cyclotron period.

The semiclassical assessment of   $\tau$ would consist in the next:
during the scattering jump from one orbit to another, in a time $\tau$,
the carriers in their orbits complete one full loop, which implies that  $\tau=T_{c}$. Therefore
the carrier involved in the scattering ends up in the same relative position
inside the final orbit as the one it started from in the initial one. The reason
for this is that the dynamics of the orbits (Landau states) is governed on average by
the position of the center of the orbit irrespective of the carrier position inside the orbit when the
scattering takes place. Then, on average, both initial and final semiclassical positions are identical in their respective orbits. Then, during the {\it flight time}, the carriers in their orbits
complete one  loop and then $\tau= T_{c}$.
If next, we substitute the result $\tau=\frac{2\pi}{w_{c}}$ in the MIRO extrema
expressions, we obtain:
\begin{equation}
\frac{w}{w_{c}} = \left(\frac{1}{4}+ n \right)
\end{equation}
for the minima and,
\begin{equation}
\frac{w}{w_{c}}  =\left(\frac{3}{4}+ n \right)
\end{equation}
for the maxima. The above expressions are exactly the
same as the ones experimentally obtained previously  by Mani et al\cite{mani2,mani4}.
Therefore, we can conclude, based on our theory, that the physical origin of the 1/4-cycle
phase shift in MIRO has to do with the harmonically swinging nature of the irradiated Landau
states  and the elastic scattering between them. Radiation frequency and carrier flight time are
the key variables ruling the process.
%harmonically swinging Landau states driven by radiation and with the
%flight time it takes the carrier to jump from orbit to another when
%it is scattered off.
%Another important result involves the left node expression that now reads like
%$\frac{w}{w_{c}}  =n$. Then, the left node of the main MIRO peak ($n=1$) marks the
%position of the resonance cyclotron: $w=w_{c}$. All of the above represent
%general results for any MIRO profile.

Nevertheless, the most important result turns out to be the final expression of
irradiated magnetoresistance where the part responsible of MIRO
can be written as:
\begin{equation}
R_{xx}\propto - A  \sin \left(2\pi\frac{w}{w_{c}}\right)
\end{equation}
This result can be described as universal since
it depends only on external variables such as radiation and
magnetic field and it is totally independent of the
type of the sample semiconductor material.

%According to the radiation-driven {\it carrier} orbit model,
%the key parameter of carriers (electrons or holes)  that has to be taken into account to tell them apart
% is the effective
%mass $m^{*}$\cite{masses}.  $m^{*}$ will mainly
%affect the amplitude of MIRO through $A$ and the phase of MIRO through
% $\tau$ (or $W_{I}$).
Finally, in a scenario where
we had two different type of carriers simultaneously coupled to
MW, as light and heavy holes, $R_{xx}$ would be written as:
\begin{equation}
 R_{xx}\propto -\left[A_{lh} \sin \left(2\pi  \frac{w}{w_{c,lh}}\right) +A_{hh}\sin \left(2\pi \frac{w}{w_{c,hh}}\right)\right]
 \end{equation}
 where $A_{lh}$ and $w_{c,lh}$ are the amplitude and cyclotron frequency for the light holes
 and $A_{hh}$ and $w_{c,hh}$ for the heavy holes.
Then,  we can predict an interference profile in the standard  MIRO, or at
least a clear distorted profile,
depending on the relative values of  the light and heavy hole
effective masses.

\begin{figure}
\centering \epsfxsize=3.5in \epsfysize=3.4in
\epsffile{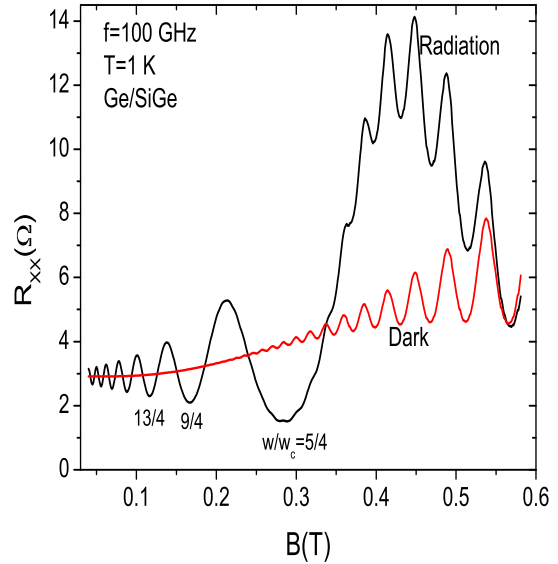}
\caption{Calculated magnetoresistance, $R_{xx}$,  vs magnetic field,  $B$, for dark and radiation
of  frequency $f=w/2\pi=100$  GHz and temperature $T=1.0K$.   For the latter
 we obtain  the radiation-induced resistance oscillations
characterized by a series of peaks and valleys in function of $B$ and
the radiation frequency $w$. The  $R_{xx}$ oscillations are calculated
for a 2DHG in a Ge/SiGe quantum well with a hole effective mass, $m^{*}=0.1m_{0}$
where $m_{0}$ is the free electron mass. Minima positions are indicated by
$w/w_{c}=5/4, 9/4, 13/4$.
}
\end{figure}

\section{Results}
In Fig. 2, we plot calculated $R_{xx}$ vs $B$ for a high mobility 2DHG hosted in a pure
Ge/SiGe quantum well corresponding to the cases of  dark and MW
of  $100$  GHz. The Ge quantum well is fully strained\cite{zudovholes,holes} and the hole density is
$p\approx 2.8 \times 10^{11} cm^{-2}$. As a result only the
heavy hole band is available to provide holes with an effective mass of $m^{*}\approx 0.1m_{0}$\cite{song}
where $m_{0}$ is the free electron mass.
For the MW curve we obtain  two oscillatory structures, one corresponds to MIRO
 characterized by a series of peaks and valleys in function of $B$ and
$w$. In other words, the standard MIRO profile very similar to the ones
previously obtained for electrons. The other structure corresponds to the Shubnikov-de Haas oscillations.
As we said above, according to our model if the hole orbits, in their oscillating
MW-driven motion, are moving backwards, on average during the
scattering jump the hole advances a larger distance than in the dark
case ( $\langle \Delta X^{MW}\rangle > \Delta X^{0}$). Therefore, we will have an increasing $R_{xx}$ and eventually a MIRO peak.
On the other hand, if the orbits are moving forwards the hole will advance on average
a shorter distance during the scattering, ( $\langle \Delta X^{MW}\rangle < \Delta X^{0}$),
giving rise to a decreasing $R_{xx}$ and a MIRO valley.
Minima positions are indicated by $w/w_{c}=5/4, 9/4, 13/4$.
As in the case of electrons, for higher $P$, one or more valleys could evolve into ZRS.
This is what we present in Fig. 3 for the same material as in Fig. 2. We obtain
a clear region of ZRS between $0.10$ and $0.12 $ T. In the inset we present
a schematic diagram explaining the physical origin of ZRS: if we increase further  the MW power,
we will eventually reach the situation  where the orbits are
moving forwards but their amplitude is larger than the scattering jump.
In this case the hole jump is blocked because the final state is occupied.

In the hypothetical case of having a 2DHG in unstrained Ge,  we would have the heavy and light hole valence bands
degenerate at the $\Gamma$ point. As a result, both type of holes would be
available to participate in the transport and  couple to MW.
The theoretical outcome, according to Eq. $17$, would be an
interference scenario that would be evidenced in $R_{xx}$.
This is presented in Fig 4 where we plot
calculated $R_{xx}$ vs $B$ for unstrained Ge, MW of frequency 100 $GHz$ and
$T=1K$. We have considered the bulk effective masses for light and
heavy holes of Ge: $m^{*}_{HH}=0.28m_{0}$ and $m^{*}_{LH}=0.044m_{0}$.
As expected, we obtain a very clear interference profile for MIRO.
For a more realistic scenario we have considered the case of a
$100{\AA}$ GaAs/GaAlAs quantum well\cite{andreani}. For this
platform it is possible, applying a uniaxial compressive stress,
to shift downwards the highest heavy hole band, making it degenerate
with light hole band at the $\Gamma$ point\cite{andreani}. The corresponding calculated
effective masses turn out to be: $m^{*}_{HH}=0.38m_{0}$ and $m^{*}_{LH}=0.09m_{0}$\cite{andreani}.
In Fig. 5 we present calculated $R_{xx}$ vs $B$ for this case.
The MW frequency is $100$ $GHz$ and $T=1K$.
 We obtain again a distorted profile for MIRO,
 revealing an apparent interference effect.

In Fig. 6 we present similar case as in Fig. 5 but now
the MW frequency is $50$ GHz and $T=0.2K$. We have lowered
the temperature in order  to weaken the damping $\gamma$, (see expression of
$A$) and obtain the corresponding resonance peaks
of  light and heavy holes. The former is observed at
$B\simeq 0.2T$ and the latter at $B\simeq 0.7T$.
Interestingly, from the $B$-position of these peaks we could obtain simultaneously
the effective masses of carriers involved in the magnetotransport.
The heavier the carrier the more displaced the peak to higher $B$.

\begin{figure}
\centering\epsfxsize=3.5in \epsfysize=3.0in
\epsffile{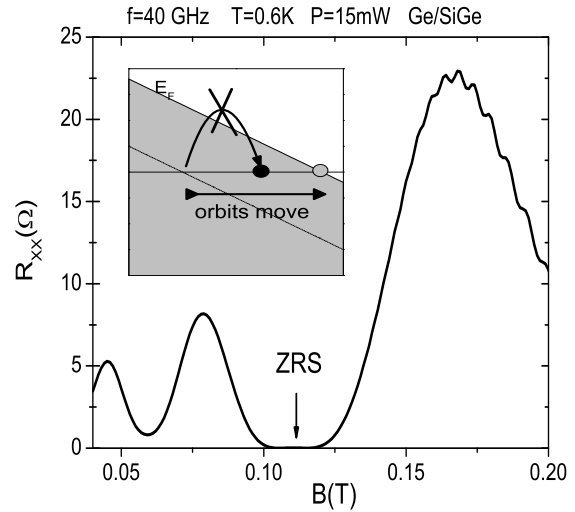}
\caption{Calculated $R_{xx}$ vs $B$ in the presence of radiation of
frequency $f=40$ GHz and $T=0.6K$. We obtain a very clear region of ZRS
between $0.10$ and $0.12 $ T. In the inset we present
a schematic diagram explaining the physical origin of ZRS. For an increasing MW power,
we will eventually reach the situation  where the orbits are
moving forwards but their amplitude is larger than the scattering jump.
In this case the hole jump is blocked because the final state is occupied.
}
\end{figure}

\begin{figure}
\centering\epsfxsize=3.5in \epsfysize=3.in
\epsffile{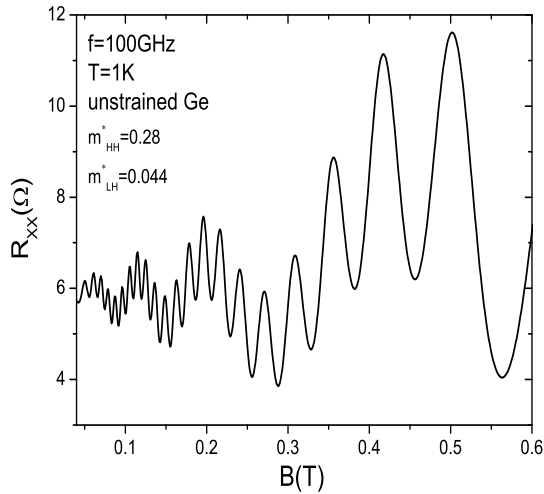}
\caption{Calculated $R_{xx}$ vs $B$ in the presence of radiation of
frequency $f=100$ $GHz$ and $T=1K$. We plot the hypothetical case of
a 2DHG in unstrained
Ge. In this case both heavy and light hole
valence band are degenerate at the $\Gamma$ point. With this scenario
we have available both types of holes to couple to MW and take part in
the magnetotransport. We observe a modulated MIRO profile as a result
of the interference regime produced by the presence of two different type
of carriers. We have considered the bulk effective masses
for Ge:  $m^{*}_{HH}=0.28m_{0}$ and $m^{*}_{LH}=0.044m_{0}$.
}
\end{figure}

\begin{figure}
\centering \epsfxsize=3.5in \epsfysize=3.0in
\epsffile{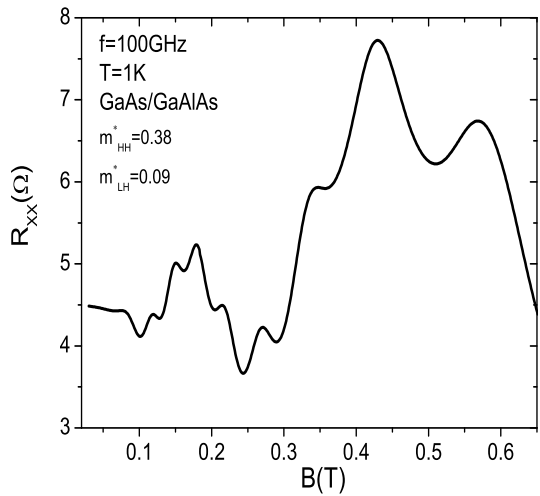}
\caption{Calculated $R_{xx}$ vs $B$ in the presence of radiation of
frequency $f=100$ GHz and $T=1K$ for a $100$ ${\AA}$ GaAs/GaAlAs quantum well.
For this
case it is possible applying a uniaxial  stress
to shift downwards the highest heavy hole band, making it degenerate
with light hole band at the $\Gamma$ point.
We obtain an interference profile for MIRO.
}
\end{figure}

\begin{figure}
\centering \epsfxsize=3.5in \epsfysize=3.0in
\epsffile{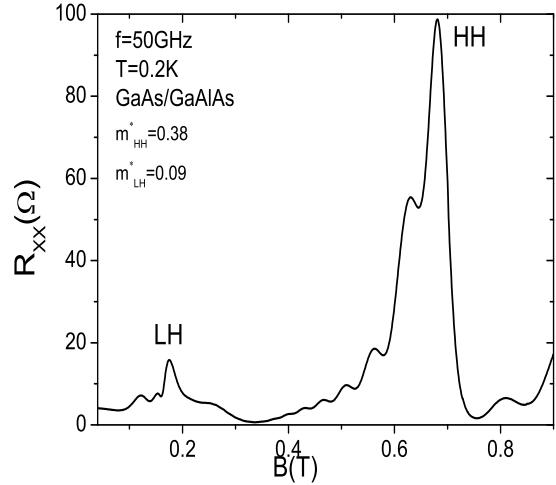}
\caption{Same as in Fig. 5 but for a MW frequency $f=50$ GHz and
$T=0.2$ K. The light hole peak  is observed at
$B\simeq 0.2T$ and the heavy hole one at $B\simeq 0.7T$.
}
\end{figure}

In Fig. 7a we present $P$-dependence of irradiated $R_{xx}$ vs $B$ for $f=40$ GHz and $T=0.6$ K
for  a 2DHG in a Ge/SiGe quantum well.
We sweep $P$ from dark to $P=10.7$ mW.
We observe that the radiation-induced oscillations
get larger as $P$ increases, showing a similar behavior as in MIRO with electrons.
In Fig. 7b, we present  $\Delta R_{xx}=R_{xx}^{MW}-R_{xx}^{0}$ vs $P$, for
data coming from the main $R_{xx}$ peak. $R_{xx}^{0}$ is the magnetoresistance
for dark and $R_{xx}^{MW}$
when the radiation field is on. We fit the data obtaining  a sublinear
$P$-dependence:
\begin{equation}
\Delta R_{xx}\propto P^{\alpha}
\end{equation}
 where $\alpha\approx0.5$ and it is straightforwardly explained with our model in terms of:
\begin{equation}
 E_{0}\propto \sqrt{P}\Rightarrow \Delta R_{xx}\propto \sqrt{P}
\end{equation}
and in agreement with previous
experimental\cite{mani6} and theoretical\cite{ina5}  results obtained for electrons.

\begin{figure}
\centering \epsfxsize=3.5in \epsfysize=5.5in
\epsffile{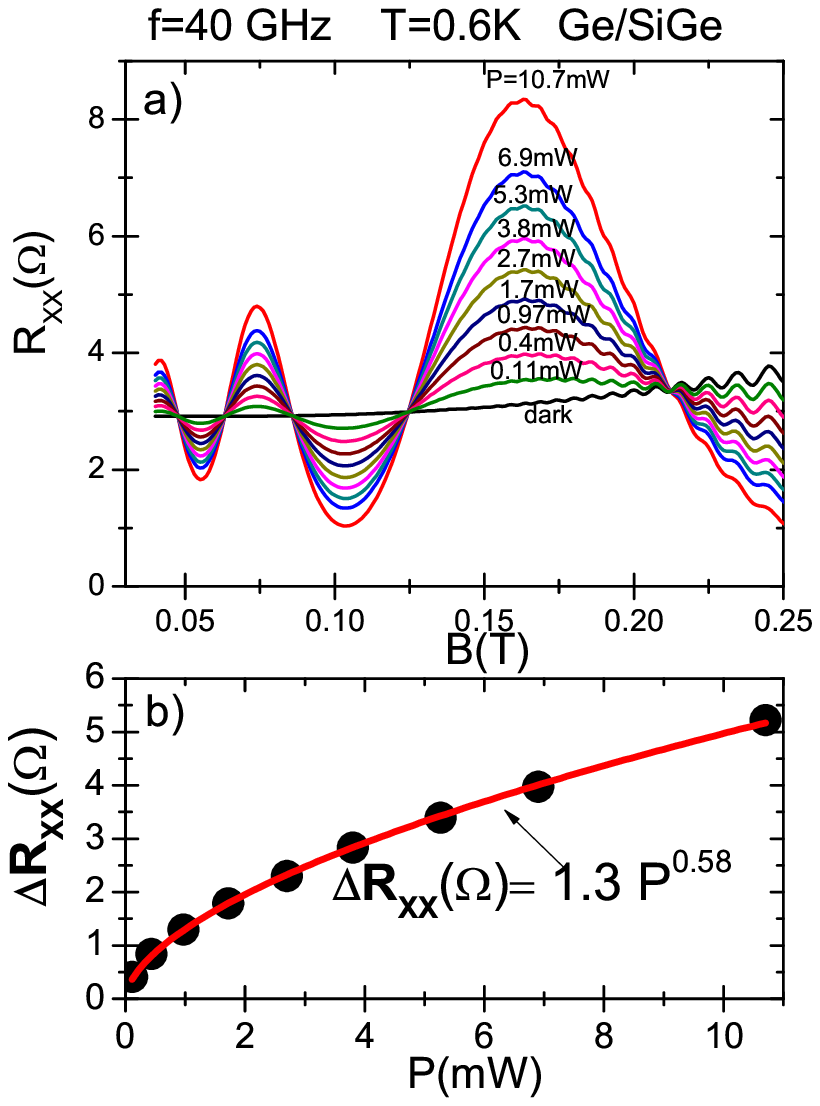}
\caption{a) Calculated $P$-dependence of $R_{xx}$ vs $B$ for $f=40$ GHz and
$T=0.6$ K for a 2DHG in a Ge/SiGe quantum well.
We sweep the radiation power $P$ from dark to
$P=10.7$ mW.
We observe that the radiation-induced oscillation
increases giving rise to larger peaks and deeper valleys.
b)  Calculated amplitude  $\Delta R_{xx}=R_{xx}^{rad}-R_{xx}^{0}$ vs $P$, for
data coming from the main peak. $R_{xx}^{0}$ is the magnetoresistance
for dark and $R_{xx}^{rad}$ when the radiation field is on. We fit the data obtaining  a sublinear
$P$-dependence where the exponent is close to $0.5$.
}
\end{figure}

\begin{figure}
\centering \epsfxsize=3.5in \epsfysize=5.5in
\epsffile{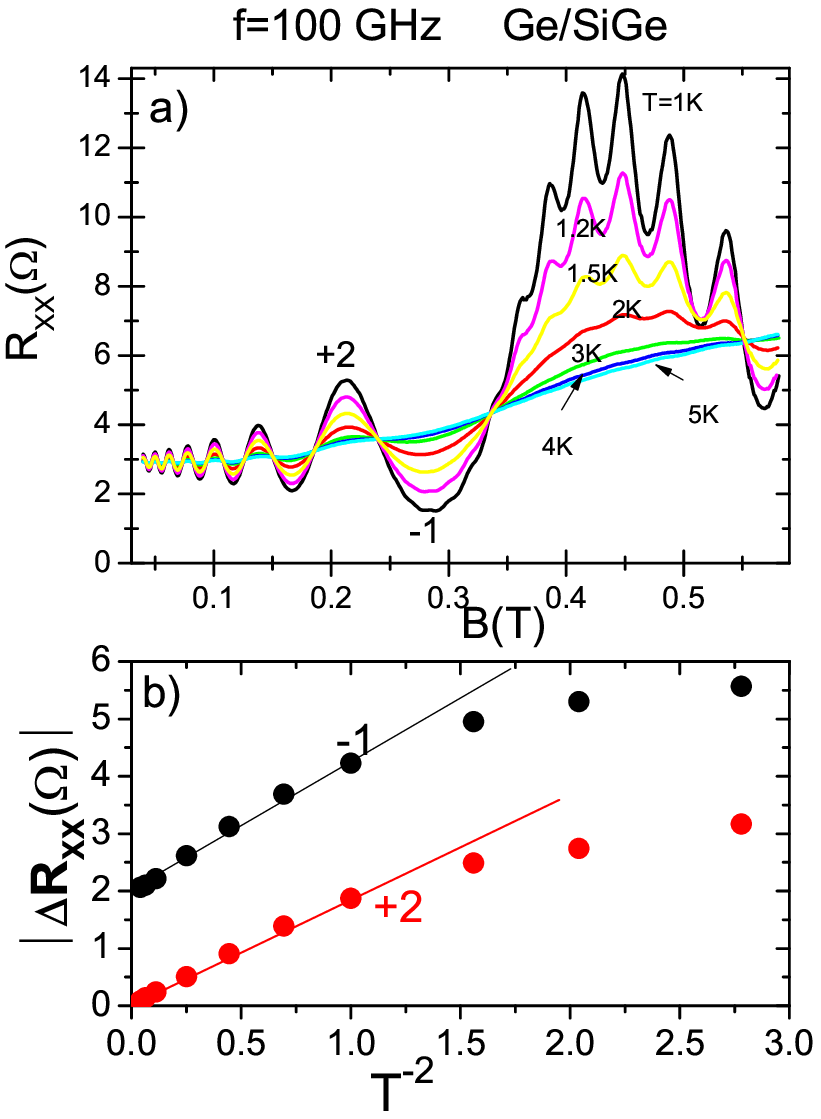}
\caption{a) Calculated $T$-dependence of $R_{xx}$ vs $B$
for $f=100$ GHz of a 2DGH in a Ge/SiGe quantum well.  We sweep the temperature from $T=1$ K
to $T=5$ K. We observe a clear decrease of the oscillation for increasing $T$,
eventually reaching a $R_{xx}$ response similar to dark, but without
the Shubnikov-de Haas oscillations that are very  affected by increasing $T$.
b) Absolute values of $R_{xx}$ amplitudes, $\Delta R_{xx}$,  of the labeled  peak $+2$
and labeled valley $-1$ vs $T^{2}$. The two curves are vertically shifted for clarity.
}
\end{figure}
In Fig 8a, we present the $T$-dependence of irradiated $R_{xx}$ vs $B$
for a MW frequency $f=100$ GHz and same material as in Fig. 7.  We sweep the temperature from $T=1.0$ K
to $T=5.0$ K. We observe a clear decrease of the oscillation amplitude for increasing $T$.
Eventually a $R_{xx}$ response similar to dark is reached, but without
the Shubnikov-de Haas oscillations that are  very affected by increasing $T$ making them
to vanish.
 The $T$-dependence, according to the model, is explained
with the damping parameter $\gamma$ which represents the interaction of carriers with the
lattice ions giving rise to  the emission of
acoustic phonons. This interaction can be calculated in terms of
the scattering rate of carriers with longitudinal acoustic phonons through the
Fermi's golden rule\cite{davies,ando}. Thus, $\gamma$ turns out to be  linear with $T$ and then an increasing $T$ means
an increasing $\gamma$ and smaller $R_{xx}$ oscillations. When the damping
is strong enough (higher $T$) $R_{xx}$ oscillations collapse.
In Fig. 8b, we present  $\Delta R_{xx}$, of the peak labeled
in the upper panel with $+2$  and the valley labeled with $-1$ vs $T^{-2}$.
The two curves are vertically shifted for clarity.
We observe that for hight $T$,  $\Delta R_{xx}$ is approximately linear with  $T^{-2}$ and for low $T$,
 $\Delta R_{xx}$ falls below the linear dependence approaching a constant value, i. e., independent
 of $T^{-2}$. We can find   explanation considering
 the expressions obtained from the model:
\begin{equation}
\Delta R_{xx}\propto \frac{e
E_{o}}{m^{*}\sqrt{(w_{c}^{2}-w^{2})^{2}+\gamma^{4}}}
%\propto \frac{e
%E_{o}}{m^{*}\sqrt{(w_{c}^{2}-w^{2})^{2}+T^{4}}}
\end{equation}
and
%\begin{equation}
$\gamma \propto T$.

%\end{equation}
Accordingly, with high $T$, $\gamma^{4}>(w_{c}^{2}-w^{2})^{2}$ and then
we can approximate $\Delta R_{xx}\propto T^{-2}$ giving a linear dependence.
Yet, for low $T$, $\gamma^{4}<(w_{c}^{2}-w^{2})^{2}$ making $\Delta R_{xx}$
independent of $T$ and approaching a horizontal line as plotted  in Fig. 8b.
To confirm this last result we have calculated $R_{xx}$ for much lower $T$ reaching
50 mK. We obtain that $\Delta R_{xx}$ tends clearly to a constant value.
We present these results in Fig 9. where we plot $\Delta R_{xx}$ versus $T$.
Here we sweep $T$ from 50 mK to 5K. According to out theory, when $T$ and in turn
$\gamma$ tend to 0, $\Delta R_{xx}$ tends to a constant value, when $w_{c}$ is
far from resonance, as can be obtained from equation [20]. This is simply a
prediction, that can be applies also to electrons, from our theoretical model because experiments on MIRO
studying T-dependence have not reached so low temperatures to date.
Therefore, the real MIRO behavior at such
very-low-temperature could serve to discriminate among
the existing theories and give credibility to the ones predicting
similar results as experiments.

\begin{figure}
\centering \epsfxsize=3.5in \epsfysize=3.0in
\epsffile{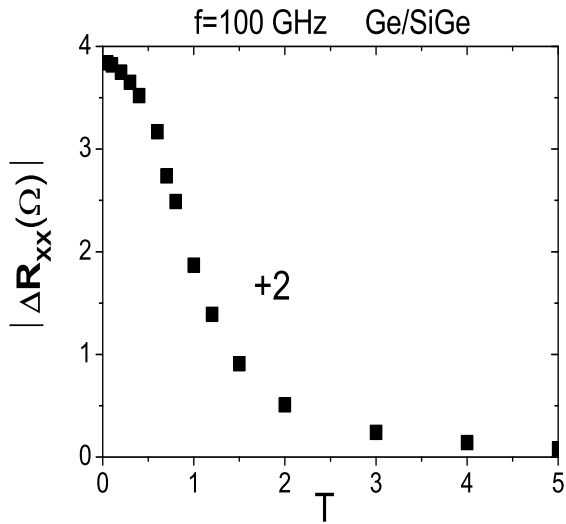}
\caption{Calculated $T$-dependence of absolute values of $R_{xx}$ amplitudes, $\Delta R_{xx}$ vs $T$
for $f=100$ GHz of a 2DGH in a Ge/SiGe quantum well.  We plot
the data corresponding to MIRO peak labelled +2. We sweep the temperature from $T=50$ mK
to $T=5$ K. We observe that $\Delta R_{xx}$ tend to a constant value as $T$ tends to 0.
}
\end{figure}

%We fit the calculated data with rational and exponential functions.
%The rational curve (red) of Fig. 4b shows the relation $R_{xx}\propto T^{-2}$ in agreement with
%the applied theoretical model in this article. The other one (blue) shows
%an exponential fit in agreement with  fits carried out in previous
%experimental results\cite{mani1,zudov1,zudov2}.
%Then, the radiation-driven
%electron orbits model could explain temperature dependence obtained in
%other experimental outcomes in the present topic\cite{mani1,zudov1,zudov2}.

\section{Conclusions}
In summary, we have presented a theoretical study on MIRO
 in a 2DHG hosted in a Ge/SiGe quantum well in order to
demonstrate
that MIRO and zero resistance states are
universal effects.
We obtain calculated  $R_{xx}$ revealing MIRO
and ZRS. We have
analytically deduced a universal expression for the irradiated magnetoresistance,
explaining the origin of the minima positions and their $1/4$ cycle phase shift.
The outcome is that these phenomena  are universal and only depend on
 radiation and cyclotron frequencies. On the other hand, they turn out to be
 independent of the type of semiconductor material.
Interestingly, we study the possibility of having simultaneously two different
carriers driven by radiation: light and heavy holes. As a result
the calculated magnetoresistance reveals an interference regime due
to the different effective masses of the two types of carriers. In the same way,
 we obtain two different resonance peaks at low enough
temperature,  corresponding to the two carriers.
Finally, we study the dependence on microwave power and temperature  obtaining
a similar behaviour as with electrons. In the power dependence we obtain a sublinear law which relates the
amplitude of the resistance oscillations and the applied power, being the exponent
approximately equal to $0.5$. For the temperature dependence we obtain, as expected, a
vanishing effect on the radiation-induced resistance oscillations
for increasing temperature. Interestingly, we also obtain that the amplitude
of MIRO tends to a constant values as temperatures tends to 0.

\section{Acknowledgments}

This work is supported by the MINECO (Spain) under grant
MAT2014-58241-P  and ITN Grant 234970 (EU).
GRUPO DE MATEMATICAS APLICADAS A LA MATERIA CONDENSADA, (UC3M),
Unidad Asociada al CSIC.

\end{document}